\documentclass[sigconf]{acmart}

\AtBeginDocument{%
  \providecommand\BibTeX{{%
    \normalfont B\kern-0.5em{\scshape i\kern-0.25em b}\kern-0.8em\TeX}}}

\setcopyright{acmcopyright}
\copyrightyear{2018}
\acmYear{2018}
\acmDOI{10.1145/1122445.1122456}

\copyrightyear{2020}
\acmYear{2020}
\setcopyright{licensedothergov}
\acmConference[ICPC '20]{28th International Conference on Program Comprehension}{October 5--6, 2020}{Seoul, Republic of Korea}
\acmBooktitle{28th International Conference on Program Comprehension (ICPC '20), October 5--6, 2020, Seoul, Republic of Korea}
\acmPrice{15.00}
\acmDOI{10.1145/3387904.3389284}
\acmISBN{978-1-4503-7958-8/20/05}

\usepackage{listings}
\begin{document}


\title{Inheritance software metrics on smart contracts }

\author{Ashish Rajendra Sai}
\email{17053145@studentmail.ul.ie}
\affiliation{%
  \city{Lero, University of Limerick}
  \state{Ireland}
}

\author{Conor Holmes}
\affiliation{%
  \institution{CSIS}
  \streetaddress{University of Limerick}
  \city{University of Limerick}
  \state{Ireland}
}

\author{Jim Buckley}
\affiliation{%
  \city{Lero, University of Limerick}
  \state{Ireland}
}

\author{Andrew Le Gear}
\affiliation{%
  \institution{Horizon Globex Ireland DAC}
  \streetaddress{Nexus Center, University of Limerick}
  \city{Limerick}
  \state{Ireland}
}

\renewcommand{\shortauthors}{Sai, et al.}

\begin{abstract}

Blockchain systems have gained substantial traction recently, partly due to the potential of decentralized immutable mediation of economic activities. Ethereum is a prominent example that has the provision for executing stateful computing scripts known as Smart Contracts. These smart contracts resemble traditional programs, but with immutability being the core differentiating factor. Given their immutability and potential high monetary value, it becomes imperative to develop high-quality smart contracts. Software metrics have traditionally been an essential tool in determining programming quality. Given the similarity between smart contracts (written in Solidity for Ethereum) and object-oriented (OO) programming, OO metrics would appear applicable. In this paper, we empirically evaluate inheritance-based metrics as applied to smart contracts. We adopt this focus because, traditionally, inheritance has been linked to a more complex codebase which we posit is not the case with Solidity based smart contracts. In this work, we evaluate the hypothesis that, due to the differences in the context of smart contracts and OO programs, it may not be appropriate to use the same interpretation of inheritance based metrics for assessment. 

\end{abstract}




\maketitle

\section{Introduction}
 Blockchain is an immutable database shared by the participants of a distributed network \cite{sai2019assessing}. The main application of blockchain, thus far, has been cryptocurrencies such as Bitcoin and Ethereum. These blockchain-based cryptocurrencies are estimated to have a market capitalization of over 200 Million USD \cite{sai2019privacy}. 
 
 Some blockchains have the provision for executing user programs in a decentralized, stateful fashion. These are known as a smart contracts \cite{buterin2014next}. Smart contracts were first implemented in Bitcoin, but were limited in functionality due to the lack of Turing completeness of the programming environment \cite{salter2016bitcoin}. This limitation was overcome with the introduction of the Turing complete execution environment in the Ethereum network. Ethereum smart contracts can be written in a high-level OO programming language known as Solidity \cite{parizi2018smart}. Solidity has a syntax that resembles that of C++, Java, C\#, and JavaScript. 

 Given the monetary value associated with smart contract transactions, quality assessment and evaluation \cite{kelly2006context} becomes vital but, even in traditional software systems, such quality assessment techniques can be quite informal \cite{ali2018architecture}. Although some argue for instant feedback regarding the quality control \cite{buckley2013jittac}, another approach is to apply well-established software metrics. Software metrics have conventionally been used to determine the quality, maintainability, and testability of programs \cite{fenton2014software}, and many programming-paradigm specific software metrics have been proposed \cite{chidamber1994metrics,ryder2005software,ryder2004software}. As Solidity adheres to the OO programming paradigm, the metrics proposed by \cite{chidamber1994metrics} have been identified as potentially useful \cite{tonelli2018smart,hegedus2019towards}. These metrics may help the developers produce improved quality contracts that are easier to test and maintain. 

Despite the syntactical and conceptual similarities between Solidity and conventional OO programming languages, they vary significantly. The immutability imposed on smart contracts by the target blockchain platform is the prime differentiating factor. Another significant difference is the cost of execution, where each byte code-level operation in Ethereum carries a dynamically calculated fee influenced by market forces within the network. Other differences include the functional limitations of Solidity.  For example, the inability to return an array from a function \cite{SolidityDocumentation,sergey2017concurrent}. Developers have found several workarounds to these limitations.  For example, where inheritance is favoured over composition \cite{stover}.  

These differences may warrant a recalibration of existing OO metrics when applied to smart contracts. In this study, we aim to evaluate two such OO metrics empirically. Given the assertion in \cite{stover}, that Solidity may favor inheritance over composition, we restrict our focus to the inheritance-based metrics proposed by \cite{chidamber1994metrics}. This early research aims to provide preliminary evidence towards answering the following research question:

\textit{Does moving from traditional OO programming contexts to Solidity contracts impact inheritance-based practice, as measured by inheritance-based metrics?}

In conventional software systems, a high degree of inheritance is often linked with greater complexity, resulting in more faults \cite{cartwright1998empirical}. It has been argued that misuse of such functionality should also be investigated \cite{amann2018systematic}. However, as functional and economic constraints limit the complexity of smart contracts, the codebase remains relatively simple. We anticipate that this relative simplicity of code might mitigate the potential complexity introduced by inheritance and may result in greater use. Additionally, as inheritance favours code reuse, we believe that developers will \textit{favour inheritance more in smart contracts than in traditional OO programs.} We address this hypothesis through an empirical analysis.

The paper makes the following contributions:
\begin{itemize}
\item We present an illustrative case study, as a motivating example, suggesting increased use of inheritance in Solidity smart contracts.
\item  We present a small empirical study that captures inheritance-metric data for prominent smart contracts on Ethereum.
\item We present a comparison between inheritance in smart contracts and traditional systems based on our provisional findings (Section 5).
\end{itemize}

\section{Software Metrics and Smart Contracts}
Software metrics are commonly used for cost estimation, quality assessment, reliability testing, security evaluation, and complexity measurement \cite{fenton2014software}. Research suggests that such metrics should be employed in the evaluation of smart contracts owing to the financial value of the transactions they mediate \cite{tonelli2018smart,hegedus2019towards}. The paper by \cite{pinna2019massive} analyzed 10000 smart contracts and reported that the software code metrics such as line of code had lower values on average with high variances when compared with standard software. However, the research in specific metrics for smart contracts is in the early stages \cite{porru2017blockchain,destefanis2018smart,tonelli2018smart,ghassemi2018reverse,8327564}.  Due to the preliminary nature of our work in this area, we restrict our focus to the inheritance-based metrics proposed by \cite{chidamber1994metrics}. 

In OO programming, inheritance is a technique by which one class or object can reuse code from another class or object \cite{johnson1988designing}. Due to the significance of inheritance in OO programming, its implication on cost, quality, and maintainability have been thoroughly examined \cite{subramanyam2003empirical,daly1996evaluating,prechelt2003controlled,unger1998impact, singh2011investigation}. A high degree of inheritance is often linked with greater complexity \cite{cartwright1998empirical}, which translates to more faults and higher costs. This adverse impact has led to the paradigm of \textit{"Composition over Inheritance"}\cite{gamma1995design}, but it is still widely used \cite{knoernschild2002java}. Thus, it becomes crucial to capture and measure different aspects of inheritance to work towards minimizing adverse effects. In this study, we restrict our investigation to the inheritance metrics proposed by \cite{chidamber1994metrics}: 
\begin{enumerate}
\item \textbf{Depth of Inheritance Tree (DIT)} calculates the maximum length of a path from a class to the root class. \cite{chidamber1994metrics} suggests that \textit{``if the depth of Inheritance tree has more than five steps, the code is too complex''}.
\item \textbf{Number of Children (NOC)} indicates the number of immediate subclasses that depend on a class in the class structure. A large number of children classes are considered favorable due to greater code reuse.
\end{enumerate}

\section{Smart Contracts and Inheritance}
Smart contracts were created in the early 1990s \cite{bonneau2015sok,szabo1997idea} to specify an enforceable agreement for a transaction. Smart contracts are programs that can be correctly executed by a network without an external trusted authority \cite{bartoletti2017empirical}. 

An Ethereum smart contract is converted into bytecode instructions \cite{frantz2016institutions} and this byte code is then appended to a block on the Ethereum Blockchain. Most Ethereum Smart Contracts are written in Solidity, a JavaScript-like programming language \cite{dannen2017introducing}.

Solidity supports multiple inheritance. It follows the guidelines of the Python language and uses C3 Linearization \cite{dannen2017introducing}, which forces a specific order in the inheritance of classes. 

Every byte code level operation in Ethereum carries an execution weight for every execution cycle. This weighted number is known as gas in the Ethereum ecosystem \cite{wood2014ethereum}.  Each unit of gas has a variable price, driven by demand on the network, and developers tend to design smart contracts that can minimize the cost associated with execution. Inheritance in Solidity plays a significant role in reducing the gas cost associated with deployment and execution. Additionally, the use of inheritance as ’aggregation’ is not a language feature of Solidity. One workaround to implementing aggregation is through inter contract communication between separately deployed contract addresses ("internal transactions") which results in an additional cost for execution \cite{stack}.

To demonstrate the impact of inheritance on the cost, we deployed two toy smart contracts on the Ethereum Testnet \cite{iyer2018ethereum} and measure gas consumption. The first toy example illustrated in Listing \ref{fig:code1} exploits inheritance for implementation and the second, functionally identical example (Listing \ref{fig:code2}), does not.

\begin{lstlisting}[language={Python}, caption={Example of Solidity contract with inheritance}, label={fig:code1},basicstyle=\footnotesize    ] contract Base { string a;
    function fit(string _a) public { a = _a; }
    function display(string name) internal view 
    returns(string) 
    { return string(abi.encodePacked(a, name));}}
contract derived is Base {
    function retrieve() public view returns(string) {
        return display("Alice");}}\end{lstlisting}

\begin{lstlisting}[language={Python}, caption={Example of Solidity contract without inheritance}, label={fig:code2},basicstyle=\footnotesize    ]
contract Display {string w;
    function Set(string _a) public {a = _a;}
    function display(string name) public view 
    returns(string) 
    {return string(abi.encodePacked(a, name));}}
contract Request { address display;
    constructor(address _display) public{display = _display;}
    function Output(string a) public 
    {Display(display).Set(a);}
    function Get() public view returns(string) 
    {return Greeting(greeting).display("Alice");}}\end{lstlisting}

The results from our cost analysis are reported in Table \ref{tab2}. Listing 1, using inheritance, yields a 54.24\% lower deployment cost and a 5.4\% lower execution cost. We expect that this significant difference in cost may impact the design choices made by Solidity developers in favor of inheritance usage.

\begin{table}
  \caption{Cost associated with inheritance}
  \label{tab2}
  \begin{tabular}{ccl}
    \toprule
    Contract Type &Deployment Cost&Execution Cost\\
    \midrule
    Inheritance (Listing 1) & 264164 gas & 42966 gas\\
    No Inheritance (Listing 2) & 575809 gas & 45412 gas\\
   \bottomrule
\end{tabular}
\end{table}

\begin{figure}[htbp]
\centerline{\includegraphics[scale=0.28]{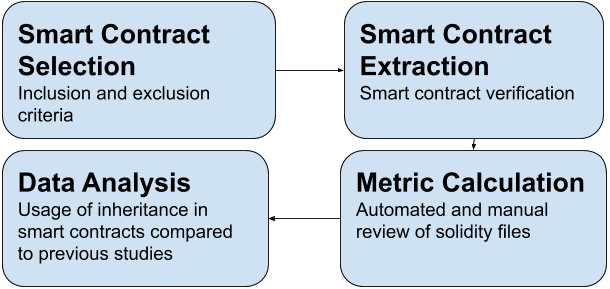}}
\caption{Study Design }
\label{fig}
\end{figure}

This example suggests two hypotheses: 
\textbf{1)} Developers will favor a high DIT to limit the \textit{cost of deployment}.
\textbf{2)} As inheritance induces functional dependencies between a child and a parent class, it may result in a high cost of execution if used extensively. To reduce the \textit{cost of execution}, developers will favor a low number of children per class resulting in a low width of the inheritance tree.
\section{Empirical Study}

Unlike traditional commercial software systems, source code for a substantial proportion of deployed smart contracts is accessible in the public domain, as trust is established through independent verification of information (transaction and/or code). However, the blockchain only provides a bytecode level source code for each deployed smart contract. Hence, we use existing public repositories to retrieve the Solidity source code. Our study design is based on the software metric evaluation guidelines put forth by \cite{fenton2014software} and we segment our study design into four phases (Figure \ref{fig}) which are now described :

\subsection{Smart Contract Selection}
Given that all records of smart contract deployment, are publicly accessible, an archive node\footnote{An archive node refers to a computing device that stores all the data present in the blockchain.} may be used to access this public data from the Ethereum network. To extract publicly available smart contracts from Ethereum, we deploy a full node. The node communicates with other nodes in the network to retrieve the history of all blocks in Ethereum. 

Once the deployed node has retrieved the history of the Ethereum blockchain, we use an open-source parser to extract data from the full node \cite{miller2019}. For the scope of our study, we are only interested in two fields present in the parsed block: the address and bytecode.

The resultant data set contained over one million smart contracts. We restrict our analysis to a specific type of smart contract: ERC-20, due to the high monetary value associated with ERC-20 tokens \cite{fenu2018ico}. ERC-20 is a standard defined by Ethereum to issue digital tokens on the platform \cite{buterin2015erc20}. We also restrict inclusion in our sample to prominently used ERC-20 smart contracts to limit the sample size in this preliminary study. We define prominence based on the market capitalization of the ERC-20 token during the observation.\footnote{The observations were conducted for a period of 24 hours on 10-01-2020. The value of market capitalization was retrieved from \cite{verify}} The restriction on the type (ERC-20) and prominence (contracts with market capitalization greater than 5000 USD) allow us to capture potentially high-quality smart contracts. After the selection of these 244 smart contracts, we extract their associated source code.

\subsection{Smart Contract Extraction}
Smart contracts are available in the public domain in the form of byte code. To extract information related to inheritance, this byte code must be converted to a high-level language such as Solidity. This conversion is traditionally done by using a decompiler to regenerate the high-level source code from byte code. However, as outlined by \cite{grech2019gigahorse}, all the decompilation approaches for Solidity are significantly limited and produce an intermediate representation, which might not be an accurate mapping to the actual source code. To overcome this limitation, we obtain verified smart contracts on Etherscan \cite{verify}. Again, due to the open nature of the blockchain platform, most smart contract developers are incentivised to publish their Solidity code online. Etherscan provides one such repository of verified smart contracts \cite{verify}. 

For our study, we were able to extract source code of 229 of 244 shortlisted smart contracts by using the Etherscan API. Once the source code of shortlisted smart contracts is successfully obtained, we perform the metric calculation.  

\subsection{Metric Calculation}

For each shortlisted smart contract, we calculate the inheritance metrics outlined in Section 2. To aid the process of metric calculation, we use the open-source static code review tool proposed by \cite{hegedus2019towards}. However, that static tool does not allow for the calculation of the number of children. We accomplish this through a manual investigation.

\subsection{Data Analysis}

\begin{figure}[htbp]
\centerline{\includegraphics[scale=0.20]{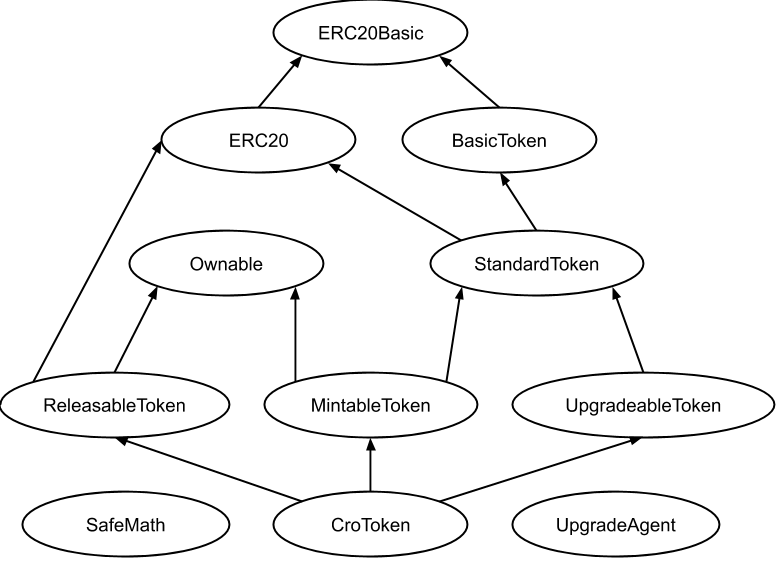}}
\caption{Inheritance Tree Diagram for CRO contract}
\label{figTwo}
\end{figure}

The 229 extracted Solidity files employed in this study are available in the project repository \footnote{Solidity source code and respective depth of inheritance graphs are available at https://anonymous.4open.science/r/a35ba37f-9e20-4073-9309-bcf8d74b75d5/.}. 
To calculate the number of children, we generate the inheritance tree diagram for all the shortlisted smart contracts. Figure \ref{figTwo} is an illustration of an inheritance tree diagram for a contract from our test sample. We can extract both the value of the DIT and the number of children from the diagram. However, a smart contract often contains multiple classes, to accommodate multiple inheritance paths per contract, we report the \textit{highest} DIT in the tree, along with an average number of children per class. 

\section{Results}

We report findings from our analysis in Table \ref{tab2}. We observe that smart contracts, on average, tend to have a DIT average of 3.29. The highest value of DIT observed in our test sample was 7, with the lowest value being 0. It should be noted that the average reported value for DIT is still below the upper threshold of 5 set by \cite{chidamber1994metrics,daly1996evaluating}.

In empirical results from \cite{tang1999empirical}, they reported an average DIT value of 1.25, 1.54, and 0.89 for systems with 5.6k,21.3k, and 16.0K lines of code, respectively. We obtained an average of 3.29 for DIT for the Solidity contracts we studied, where the contracts were, on average, 328 LOC long. This suggests a preference towards using inheritance in smart contracts when compared to conventional programming languages, particularly when normalized for LOC. Despite the overall trend towards favoring inheritance, we report that the lowest value of DIT observed for our data sample is 0, which, according to \cite{chidamber1994metrics}, violates the OO paradigm. This implies that the lower bound on DIT values set by \cite{chidamber1994metrics} is not suitable for all smart contracts, as inheritance is not always required in order to design an executable smart contract: 4\% of smart contracts in our data sample reported a DIT value of 0. 

\begin{table}
  \caption{Inheritance Metrics for Smart Contracts}
  \label{tab2}
  \begin{tabular}{cccl}
    \toprule
    Metric & Average & Median & Standard Deviation\\
    \midrule
    Depth of Inheritance & 3.29 & 3.6 & 1.40\\
    Number of Children & 0.99 & 1.0 & 0.45\\
   \bottomrule
\end{tabular}
\end{table}

The higher average inheritance finding points to differences between traditional OO software and smart contracts: possibly because of the cost associated with smart contract execution, as presented in Section 2. Surprisingly, the average number of children (NOC) for our data sample was 0.99, which seems to oppose the recommendation of \cite{chidamber1994metrics}, where they advocate for a higher number of children, for more code reuse. We speculate that the execution cost of employing extensive code reuse may be the reason behind a low number of children per class, as argued in hypothesis 2. Code reuse is usually achieved by calls to existing code, which in smart contracts may result in higher cost, as every call carries an execution cost. It may be more cost-efficient to rewrite the piece of code rather than employing extensive code calls.  In the case of NOC, the trade-off is between a high degree of code reuse, i.e., a high value of NOC, and a lower cost of execution, which, overall, seems to result in the observed low value of NOC. Our study reports that the smart contract with the highest number of children per class had a value of 2.12. This is quite low in comparison to the highest number of children per class of 16, as reported  in \cite{tang1999empirical}.

\section{Conclusion}
Measurement of quality, reliability, and security are some of the application areas of software metrics \cite{fenton2014software}. As software systems tend to vary significantly depending on development and deployment environment, a single interpretation of a suite of metrics may not apply to all types of software systems.

In this paper, we performed an evaluation of prominent smart contracts to assess the applicability of traditional inheritance based software metrics in a smart contract context (Section 4). We observe that the metrics for this environment do not equate to the same metrics for their more traditional environment, suggesting that factors related to context are at play: For example, the smaller source code size of the Ethereum smart contract. This smaller size may be a result of the functional limitations imposed by the decentralized nature of the Ethereum platform and, along with a significant difference in cost associated with execution, may result in different interpretations of, and different values for that metric suite.

Results from the static and manual code investigation reported a trend towards a higher depth of inheritance in examined smart contracts in accordance with hypothesis 1. But we also note that the observed average value DIT (2) still falls below the higher threshold of 5 proposed by \cite{chidamber1994metrics}. We argue that due to the less complex nature of smart contracts, and the lower cost associated with inheritance, the reported values of DIT may signal a deliberate design choice of developers. We also analyzed and reported the average number of children per smart contract. The results suggest a lower NOC in accordance with our hypothesis 2, arguably due to the cost of execution associated with more child classes, as illustrated by the toy example in Section 2. 

As the execution of these smart contracts attracts an execution cost known as a transaction fee, the design decisions may be driven by the need to minimize the cost. We also argue that the economics of smart code execution may drive the choice of preferring inheritance over alternate design choices such as composition. The opportunity cost associated with the higher width of inheritance (more child per class) may also be the reason behind the low value of NOC. Hence, we argue that a direct application of traditional OO metrics to smart contracts may not yield useful results. The notion of complexity and inheritance is significantly different in smart contracts due to the resource constraints of the execution environment.  

However, the study presented in this ERA paper is limited by the choice of examined smart contracts. We have only studied a small sample of contracts. Our argument against the use of traditional inheritance metrics is based on the results from this sample, which may be biased based on the prominent use-case of smart contracts. That is, the analyzed smart contracts are sampled from pool of ERC-20 contracts and share a significant degree of similarity. We also acknowledge that the static parser and manual analysis phase my suffer from false positives. We intend to conduct the study on a larger data-set in the future. We also aim to expand our metric evaluation to other metrics. Examining more traditional metrics for smart contracts may guide new thresholds for  metrics specific to smart contract context, and prompt new metrics. A further investigation into this field may prove to be vital for the smart contract development. We also intend to examine security metrics for smart contracts. The security of smart contracts has been extensively explored \cite{lin2017survey} and can be used as a measure of quality.

\begin{acks}
This work was supported, in part, by Science Foundation Ireland grant 13/RC/2094 
\end{acks}

\bibliographystyle{ACM-Reference-Format}
\bibliography{sample-base}


\end{document}